\documentclass[12pt, a4paper]{article}
\usepackage{lineno}
\usepackage{color}
\usepackage{graphicx}
\usepackage[toc,page]{appendix}
\usepackage{caption}
\usepackage{amsmath}
\usepackage{subcaption}
\usepackage{hyperref}
\usepackage{mathtools}
\usepackage{float}
\usepackage{rotating}
\usepackage{url}
\usepackage{enumitem}
\usepackage[table]{xcolor}
\usepackage{multirow}
\usepackage[font=scriptsize,labelfont=bf,skip=6pt]{caption}
\usepackage[T1]{fontenc}
\usepackage{authblk}
\font\myfont=cmr12 at 15pt

\providecommand{\keywords}[1]{\textbf{\textit{Key Words--}} #1}

\begin{document}
\title{\myfont Impact of Single-Mask Hole Asymmetry on the Properties of GEM Detectors}
\author{Aashaq Shah\thanks{aashaq.shah@cern.ch}}
\author{Ashok Kumar}
\author{Md. Naimuddin}
\affil{Department of Physics $\&$ Astrophysics, University of Delhi, Delhi, India}
\author{Archana Sharma}
\author{Jeremie Alexander Merlin}
\affil{CERN, Geneva, Switzerland}
\maketitle
\begin{abstract}
A single-mask Gas Electron Multiplier (GEM) technique overcomes the cumbersome practice of alignment of two masks and allows the production of foils with very large area as needed for the CMS muon forward region upgrade. However, the holes obtained with refinements in the single-mask technique are asymmetrically bi-conical in shape compared to symmetrically bi-conical holes of double-mask technology. The hole geometry and their uniformity define the performance of the detectors which are constructed with such GEM foils. To evaluate the effect of this asymmetry, the foils have been characterized experimentally using a special prototype with three single-mask GEM foils. The structure allowed to change the orientation of foils, testing from above with foils having a large hole opening, testing from the bottom  with all the foils having small hole opening. The effective gain, energy resolution and the charging up behavior are compared for the two different hole orientations.
\end{abstract}
\section{Introduction}
The instantaneous luminosity at the Large Hadron Collider (LHC) will exceed 2 $\times$ 10$^{34}$ cm$^{-2}$s$^{-1}$ after the second Long Shutdown (LS2) upgrade expected in 2019-2020. An additional set of muon detectors known as GE1$\slash$1 will be installed during LS2 in the first end-cap muon station in the region  $1.6<\left|\eta\right|<2.2$ of the CMS experiment \cite{one, two} to enhance the muon trigger and reconstruction capabilities. The proposed GE1$\slash$1 station uses gas electron multiplier (GEM) technology \cite{three} which has a high rate capability and is an excellent choice as a muon detector in the high $\eta$ region of the CMS experiment \cite{four}. The GE1$\slash$1 upgrade paves the way for further CMS muon system upgrades with GEM detectors for the High-Luminosity phase of the LHC.
\section{Test Setup}
\begin{figure}[ht]
\centering
\includegraphics[width=2.5cm, height=2 cm]{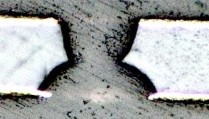}
\includegraphics[width=3.4cm, height= 2.5cm]{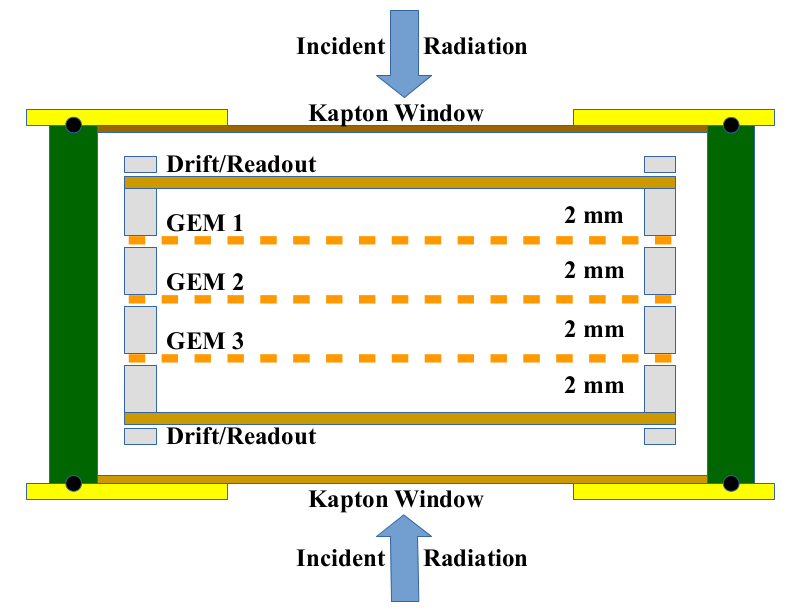}
\caption{Scanning Electron Microscope (SEM) image showing the asymmetric single hole structure in GEM produced by single-mask technique (left) and a sketch of the tripple GEM used for the meassurements (right).}
\label{fig:HoleStructure}
\end{figure}
The GE1$\slash$1 upgrade of the CMS experiment at the LHC CERN requires large area GEM foils. These GEM foils are produced by the single-mask technique \cite{five}. However, single-mask technique gives rise foils with asymmetric holes with a wide hole opening (85 $\mu$m) on one side and slightly a narrow hole opening (70 $\mu$m) on the other side as depicted in Figure~\ref{fig:HoleStructure}. We irradiate both these openings, in one case wide opening faces the source and in the other case, narrow opening facing the source. For the measurement purposes we name first case as `Orientation A' and later case as `Orientation B' respectively. Therefore, to study the effect due to this asymmetry on the properties of the detector, a special detector with standard 10 cm $\times$ 10 cm GEM foils with a symmetric gap configuration of (2$\slash$2$\slash$2$\slash$2) mm for drift, transfer-1, transfer-2, and induction gap was constructed with the provision that the detector could be irradiated on both the directions i.e from the top and from the bottom side with the appropriate applied fields by inverting the external resistive divider used to provide the potentials across the foils and gaps (Figure~\ref{fig:HoleStructure}).  This allows to irradiate the GEM foils with either `Orientation A' or with `Orientation B' without opening the detector to change the orientation of the foils and hence  minimizes the errors and interference's that may occur due to opening and closing of the detector. 
\begin{figure}[ht]
\centering
\includegraphics[width=5.7cm, height=4.6cm]{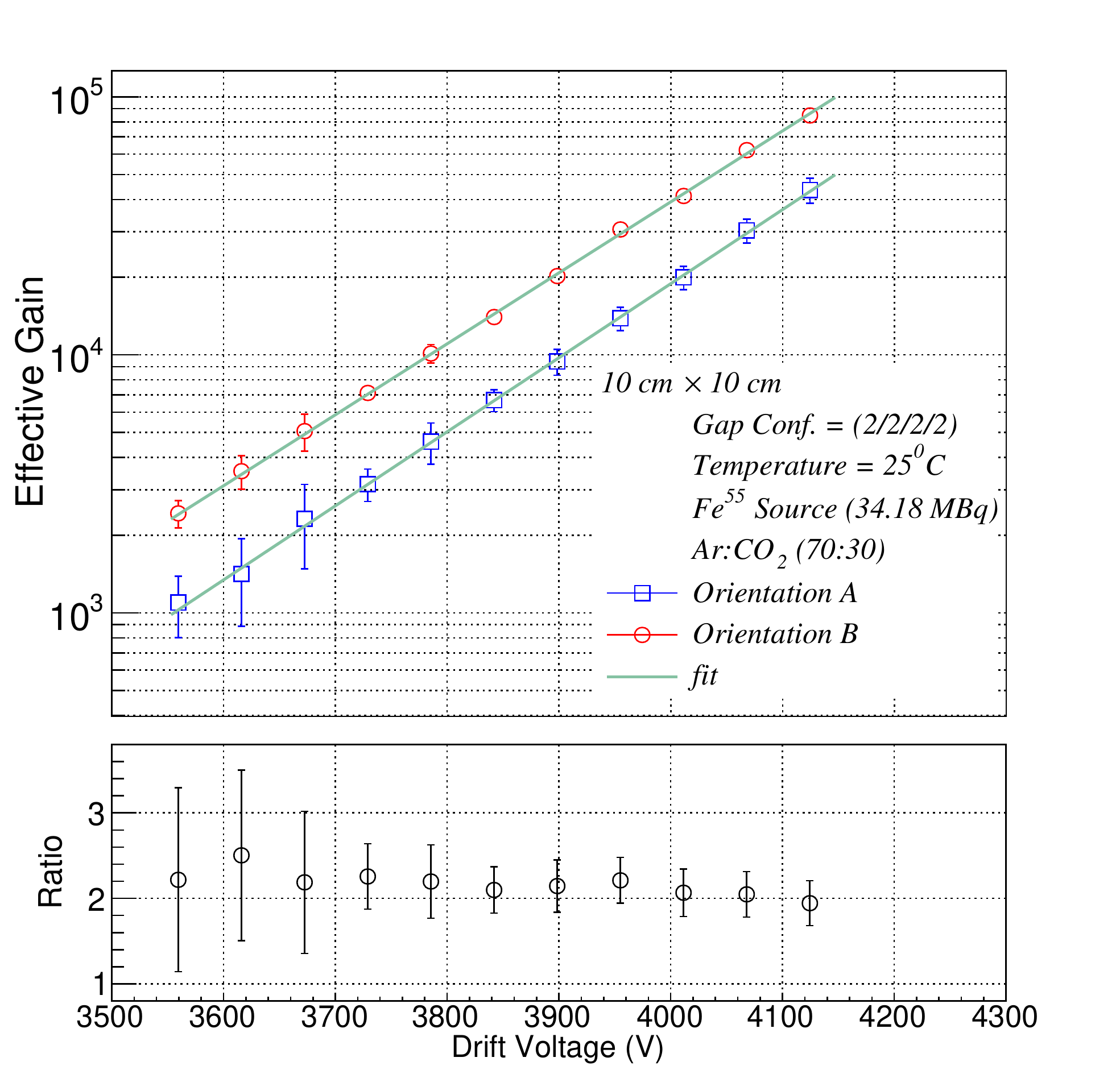}
\caption{Gain measured in a triple GEM 10 cm $\times$ 10 cm detector when all the three foils either with `Orientation A' or with `Orientation B' were facing the incident source (Fe$^{55}$). The ratio plot in the bottom shows that the gain is almost 2 times higher in `Orientation B' compared to `Orientation A'.}
\label{fig:Gain}
\end{figure}
\section{Gain Measurements} \label{sect:GainMeasurements}
The detector was irradiated by a $^{55}$Fe source producing photons with an average energy of 5.9 keV that are fully converted in electron-ion pairs in the drift gap via the photoelectric effect in the gas of the drift region. The amplified current I$_{a}$ induced on the bottom of the third GEM foil was measured with a KEITHLEY 6487 pico-ammeter connected to the anode and the effective gain "G" was estimated. The effective gain determined for `Orientation A' and `Orientation B' is shown in the Figure~\ref{fig:Gain}. The gain is observed to be higher in `Orientation B' compared to `Orientation A' by a factor of almost 2. 

\begin{figure}[ht]
\centering
\includegraphics[width=5.7cm, height=4.6cm]{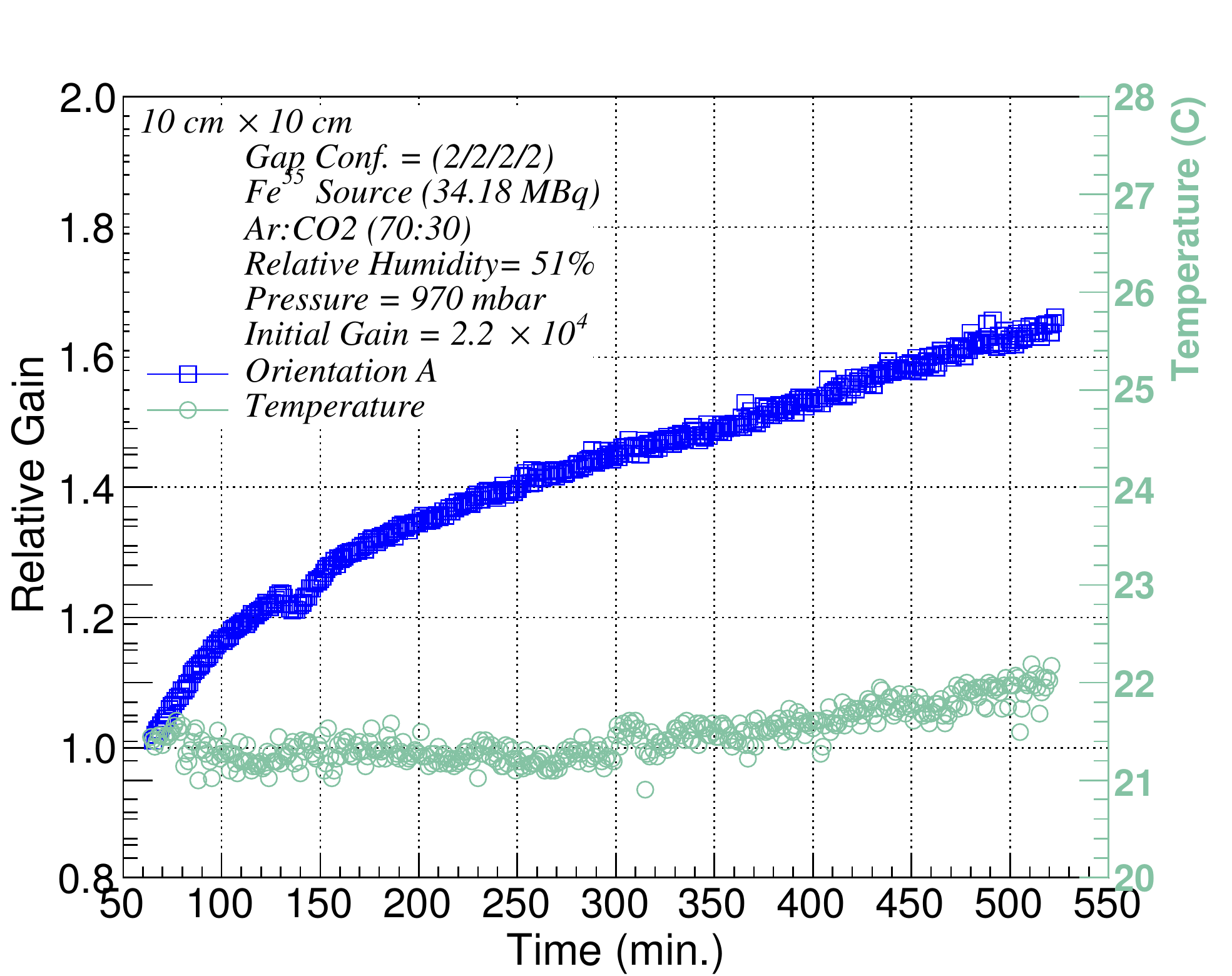}
\caption{Normalized gain as a function of time for `Orientation A' when detector was under continuous irradiation. Temperature has also been plotted to demonstrate the stability of the ambient conditions, while humidity and pressure were noticed to be constant during the period of measurements}
\label{fig:ChargingUp}
\end{figure}
\section{Charging Up} \label{sect:ChargingUp}
Effective gain as a function of time when the detector was under continuous irradiation with a source has also been determined for both the orientations. While the gain increases continuously in `Orientation A' upto 1.6 times the initial gain of 2.2 $\times$ 10$^{4}$ during the period of measurement as shown in Figure~\ref {fig:ChargingUp}, while in `Orientation B' gain is observed to be almost flat with respect to the initial value. The observed differnces in gain Section~\ref{sect:GainMeasurements} and the charging up behavior may be attributed to avalanche multiplication within the GEM holes in which charges deposit on the insulator and dynamically modify the gain, an effect that depends upon the geometry of the hole. The gain increases with time and exposure to radiation, an observation that can qualitatively be understood as due to the increase of the electric field within the hole induced by accumulating charges on the walls, opposing additional depositions on the surface~\cite{RD_51}. 

\section{Energy Resolution}
\begin{figure}[ht]
\centering
\includegraphics[width=5.7cm, height=4.6cm]{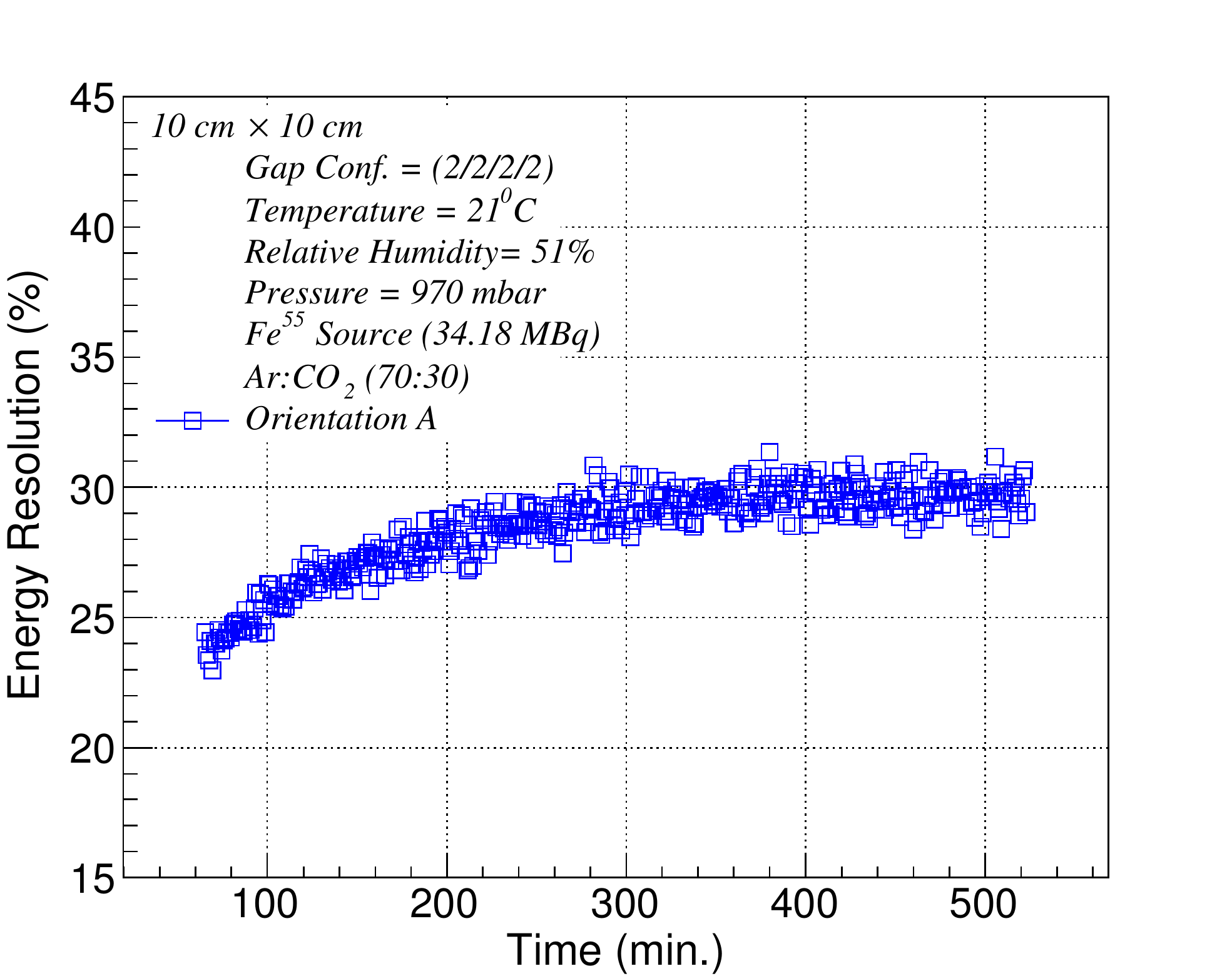}
\caption{Energy resolution determined as a function of time for `Orientation A' with the initial gain of 2.2 $\times$10$^{4}$ when detector was continuously irradiated by Fe$^{55}$ source. }
\label{fig:ResolutionTime}
\end{figure}

The energy resolutions were determined using $^{55}$Fe source at an equal initial gain of 2.2 $\times$ 10$^{4}$ and the best obtained values are ~23.71$\%$$\pm$0.02 and ~18.06$\%$ $\pm$0.01 for `Orientation A' and `Orientation B', respectively. Further, the resolution's were determined when the detector was continuously irradiated by $^{55}$Fe source over a period of time. The resolution in `Orientation A' worsens with respect to the initial value of 24$\%$ and becomes nearly flat at 30$\%$ as shown in Figure~\ref{fig:ResolutionTime}, the effect might be attributed to the charging up behavior as noted in Section~\ref{sect:ChargingUp}. On the other hand the initial value of the resolution in `Orientation B' has been observed to be 18$\%$ and stabilizes at around 20$\%$.
\section{Results and Outlook}
Single-mask foils with asymmetric holes were tested for gain, resolution, charging up etc. While some of the similar results like gain and rate capability measured in some different test campaign were reported by CMS Muon group in MPGD \cite{six} and we confirm the results. We observed that the hole asymmetry strongly affects the properties of the detector and the `Orientation B' facing the incident radiation performs better compared to the `Orientation A'.
%
%
%
%
%

\end{document}